\documentclass[a4paper,11pt]{article}
\usepackage{aaskaiid}
\usepackage{orcidlink}
\usepackage{comment}
\usepackage{url}

\usepackage{xcolor} 

\title{Probing Physical Conditions in Classical and Symbiotic Novae with the Square Kilometre Array Observatory}
\ShortTitle{Nova science in the era of the SKAO}

\author[1,2]{Rocco Lico \orcidlink{0000-0001-7361-2460}}
\ShortName{Lico, Giroletti et al. 2026} 
\author[1]{Marcello Giroletti \orcidlink{0000-0002-8657-8852}}
\author[3]{Arnau Aguasca-Cabot \orcidlink{0000-0001-8816-4920}}
\author[4]{Joris Kersten \orcidlink{0000-0002-9121-5749}}
\author[5,6]{Benito Marcote \orcidlink{0000-0001-9814-2354}}
\author[7]{Ulisse Munari \orcidlink{0000-0001-6805-9664}}
\author[5]{Miriam M.~Nyamai \orcidlink{0000-0002-8973-4072}}
\author[8]{Timothy J.~O'Brien \orcidlink{0000-0001-7687-3543}}
\author[5]{Zsolt Paragi \orcidlink{0000-0002-5195-335X}}
\author[9,10]{Iris de Ruiter \orcidlink{0000-0002-4752-5467}}
\author[11]{Kirill Sokolovsky \orcidlink{0000-0001-5991-6863}}
\author[]{David R.~A.~Williams\textsuperscript{9} \orcidlink{0000-0001-7361-0246} on behalf of the SKAO Transients SWG.}

\affiliation[1]{INAF Istituto di Radioastronomia, via Gobetti 101, 40129 Bologna, Italy.}
\emailAdd{rocco.lico@inaf.it}
\affiliation[2]{Instituto de Astrof\'{\i}sica de Andaluc\'{\i}a-CSIC, Glorieta de la Astronom\'{\i}a s/n, 18008 Granada, Spain.}
\emailAdd{marcello.giroletti@inaf.it}
\affiliation[3]{Departament de F\'\i sica Qu\`antica i Astrof\'\i sica, Institut de Ci\`encies del Cosmos, Universitat de Barcelona, IEEC-UB, Mart\'\i\ i Franqu\`es, 1, 08028, Barcelona, Spain.}
\affiliation[4]{Department of Astrophysics/IMAPP, Radboud University, P.O. Box 9010, 6500 GL Nijmegen, The Netherlands}
\affiliation[5]{Joint Institute for VLBI ERIC, Oude Hoogeveensedijk 4, 7991 PD Dwingeloo, The Netherlands.}
\affiliation[6]{ASTRON, Oude Hoogeveensedijk 4, 7991 PD Dwingeloo, The Netherlands.}
\affiliation[7]{INAF Osservatorio Astronomico di Padova, 36012 Asiago, VI, Italy.}
\affiliation[8]{Jodrell Bank Centre for Astrophysics, School of Physics and Astronomy, University of Manchester, Manchester M13 9PL,UK.}
\affiliation[9]{Sydney Institute for Astronomy, School of Physics, The University of Sydney, NSW 2006, Australia}
\affiliation[10]{ARC Centre of Excellence for Gravitational Wave Discovery (OzGrav), Hawthorn, VIC 3122, Australia}
\affiliation[11]{Department of Physics and Astronomy, Texas Tech University, Box~41051, Lubbock, TX 79409, USA.}

\abstract{Cataclysmic variables and symbiotic stars are interacting binary systems in which a hot white dwarf (WD) orbits a companion main-sequence or red giant star, respectively. Accumulation of hydrogen-rich material on the WD surface may trigger re-ignition of thermonuclear reactions that, under degenerate conditions, lead to an explosive ejection of the accreted layer mixed with WD material. These explosions, known as classical novae, provide opportunities to study key astrophysical processes such as binary evolution, accretion, ionisation of circumstellar material, mass ejection, jet formation, and thermonuclear burning. Radio emission in these systems arises from both thermal and non-thermal processes, which manifest differently in classical and symbiotic novae. $\gamma$-ray emission has also been detected in several cases, and recent progress, driven by coordinated radio and multiwavelength observations, has greatly advanced our understanding of both types of novae.
Multi-frequency, multi-epoch, and multi-scale interferometric observations are powerful probes of the evolving physical conditions following thermonuclear explosions, revealing information from both ionised and relativistic particle populations. The SKAO, particularly its SKA-Mid component, will enable regular monitoring of Galactic novae, multiple times per year for classical novae and every few years for symbiotic systems. It will explore a wide range of conditions, including companion types, WD masses, accretion regimes, and surrounding environments. The VLBI capabilities of the SKAO will target compact shocked regions, while its exceptional sensitivity will also allow characterisation of emission during quiescent phases of the binaries.
}

\begin{document}
\newcommand{\actaa}{Acta Astron.} 
\newcommand{\araa}{Annu. Rev. Astron. Astrophys.} 
\newcommand{\aar}{Astron. Astrophys. Rev.} 
\newcommand{\ab}{Astrobiol.} 
\newcommand{\aj}{Astron. J.} 
\newcommand{\apj}{Astrophys. J.} 
\newcommand{\apjl}{Astrophys. J. Lett.} 
\newcommand{\apjs}{Astrophys. J. Suppl. Ser.} 
\newcommand{\ao}{Appl. Opt.} 
\newcommand{\apss}{Astrophys. Space Sci.} 
\newcommand{\aap}{Astron. Astrophys.} 
\newcommand{\aapr}{Astron. Astrophys. Rev.} 
\newcommand{\aaps}{Astron. Astrophys. Suppl.} 
\newcommand{\baas}{Bull. Am. Astron. Soc.} 
\newcommand{\caa}{Chinese Astron. Astrophys.} 
\newcommand{\cjaa}{Chinese J. Astron. Astrophys.} 
\newcommand{\cqg}{Class. Quantum Gravity} 
\newcommand{\gal}{Galaxies} 
\newcommand{\gca}{Geochim. Cosmochim. Acta} 
\newcommand{\icarus}{Icarus} 
\newcommand{\jcap}{J. Cosmol. Astropart. Phys.} 
\newcommand{\jgr}{J. Geophys. Res.} 
\newcommand{\jgrp}{J. Geophys. Res.: Planets} 
\newcommand{\jqsrt}{J. Quant. Spectrosc. Radiat. Transf.} 
\newcommand{\memsai}{Mem. Soc. Astron. Italiana} 
\newcommand{\mnras}{Mon. Not. R. Astron. Soc.} 
\newcommand{\nat}{Nature} 
\newcommand{\nastro}{Nat. Astron.} 
\newcommand{\ncomms}{Nat. Commun.} 
\newcommand{\nphys}{Nat. Phys.} 
\newcommand{\na}{New Astron.} 
\newcommand{\nar}{New Astron. Rev.} 
\newcommand{\physrep}{Phys. Rep.} 
\newcommand{\pra}{Phys. Rev. A} 
\newcommand{\prb}{Phys. Rev. B} 
\newcommand{\prc}{Phys. Rev. C} 
\newcommand{\prd}{Phys. Rev. D} 
\newcommand{\pre}{Phys. Rev. E} 
\newcommand{\prl}{Phys. Rev. Lett.} 
\newcommand{\psj}{Planet. Sci. J.} 
\newcommand{\planss}{Planet. Space Sci.} 
\newcommand{\pnas}{Proc. Natl Acad. Sci. USA} 
\newcommand{\procspie}{Proc. SPIE} 
\newcommand{\pasa}{Publ. Astron. Soc. Aust.} 
\newcommand{\pasj}{Publ. Astron. Soc. Jpn} 
\newcommand{\pasp}{Publ. Astron. Soc. Pac.} 
\newcommand{\rmxaa}{Rev. Mexicana Astron. Astrofis.} 
\newcommand{\sci}{Science} 
\newcommand{\sciadv}{Sci. Adv.} 
\newcommand{\solphys}{Sol. Phys.} 
\newcommand{\sovast}{Soviet Ast.} 
\newcommand{\ssr}{Space Sci. Rev.} 
\newcommand{\uni}{Universe} 

\maketitle

\section{Science background}
Interacting binaries that host accreting white dwarfs (WD) produce a remarkably rich variety of explosive and variable phenomena \citep{1986syst.book.....K,1995cvs..book.....W}.
Among these, nova eruptions, thermonuclear runaways on the WD surface \citep{2016PASP..128e1001S}, are the most dramatic ones.
They provide exceptional laboratories for studying mass transfer, 
thermonuclear burning under degenerate conditions, shock formation, particle acceleration, 
the shaping of the circumstellar environment and chemical evolution of the
Galaxy \citep[e.g.,][]{2020ApJ...895...70S,2020A&ARv..28....3D,2021ARA&A..59..391C}. 
Over the past two decades, radio astronomy has evolved from episodic single-target observations to systematic, 
multi-frequency programs that have unveiled the geometry of nova ejecta, the balance between thermal and non-thermal emission, 
and the evolution of shocks and circumstellar interactions at spatial scales inaccessible at other wavelengths
\citep[see eg.][]{2021ApJS..257...49C}. 

Depending on the nature of the companion star, novae are broadly classified as classical or symbiotic. Classical novae occur in close binaries where a WD accretes hydrogen-rich material from a Roche-lobe-filling, low-mass main-sequence or slightly evolved companion. When the accreted envelope reaches critical pressure, it ignites a runaway fusion reaction, ejecting material and brightening the system by about ten magnitudes for weeks to months \citep{2010AJ....140...34S,2021ApJ...910..120K}. After the outburst, accretion resumes leading to a new eruption on timescales from decades to thousands of years depending primarily on the mass of the WD and the accretion rate. Classical novae cover a wide range of ejected masses ($\sim10^{-7} - 10^{-4} M_\odot$) and kinetic energies ($10^{43} - 10^{47}$ erg), producing rich spectroscopic evolution and luminous thermal radio emission from their ionized, expanding ejecta.
About 10 novae are discovered in the Galaxy each year, but many more are expected to evade detection due to dust obscuration, with the total
number estimated to be around 30 \citep{2017ApJ...834..196S,2022ApJ...936..117R}
or 45 \citep{2021ApJ...912...19D,2023MNRAS.523.3555Z} per year.


Symbiotic novae are rarer eruptions occurring in binaries where a WD accretes material from a red giant (RG), either via Roche-lobe overflow or wind accretion. Owing to the presence of a RG companion, these systems have wider orbital separations, typically 
a few astronomical units, and longer orbital periods of a few years \citep{Munari2025a}. 
The dense, extended circumstellar medium profoundly alters the eruption dynamics: when the thermonuclear runaway occurs, the fast nova ejecta interact with the pre-existing, slower-moving RG wind, generating strong shocks and sustained emission from radio to $\gamma$-ray energies. This contrast in scale and ambient density between symbiotic and classical novae provides natural laboratory for studying how ambient medium and binary separation control shock strength, particle acceleration, and radiative evolution \citep[see eg.][]{li2017nova, azzollini2023multi}. 
The current census of symbiotic stars in the Milky Way includes 284 confirmed systems \citep{Merc2025}. Particularly relevant are recurrent symbiotic novae, such as RS Ophiuchi (RS\,Oph), which provide a rare opportunity to observe and compare multiple eruptions within a human lifetime at different orbital phases, varying the degree to which the red giant wind obscures our view of the erupting white dwarf \citep{Obrien2006}. Because of their short outburst recurrence times, these systems are thought to host very massive WDs near the Chandrasekhar limit, making them strong candidates as progenitors of Type Ia supernovae \citep{kato2012}.

Radio wavelengths probe both free-free emission from ionized gas and synchrotron emission from shock-accelerated particles \citep{kantharia2007giant,eyres2009double, de2023low}, providing insights into ejecta
mass, magnetic fields, and shocks. 
High-resolution, multi-frequency and multi-epoch radio observations allow us to constrain the ejecta morphology, 
asymmetries and evolution, as well as the mass ejection history, the particle acceleration mechanisms and 
the role of the circumstellar medium in shaping the remnant \citep{Obrien2006, rupen2008expanding, sokoloski2006combination, Lico2024, Munari2022}. These diagnostics were pivotal in a number 
of recent discoveries: clear evidence of bipolar and non-spherical ejecta, challenging earlier assumptions of spherical symmetry 
\citep{Obrien2006, Lico2024}, unambiguous detection of non-thermal emission associated with shocks
\citep[e.g.,][]{Munari2022,2023MNRAS.521.5453S,de2023low}, 
milliarcsecond scales complex structure connected to the pre-outburst activity, as well as the association of novae with 
$\gamma$-ray \citep{Abdo2010, cheung2022fermi} and very-high energy \citep{2022Sci...376...77H} emission, implying efficient particle acceleration \citep{Giroletti2020}.

The Square Kilometre Array Observatory (SKAO), with its unprecedented sensitivity, angular resolution, and survey capabilities  will be a game-changer in the context of transient phenomena, particularly for nova science. 
Compared with current facilities, SKA will detect much fainter thermal radio emission, produce dense multi-epoch spectral monitoring, and conduct large-scale surveys of the Milky Way, 
providing a statistically complete picture of the nova population. 
Moreover, thanks to the ability to phase-up many dishes together for participation in very long baseline interferometry (VLBI) networks, it will map the complex geometries of nova ejecta with unprecedented detail \citep{Obrien2015}. 
These capabilities (detailed in Section \ref{sec:role_of_SKA}) address the key open questions in nova physics: 
how mass is ejected (mass, velocity and geometry), how shocks form and accelerate particles, what fraction of the ejecta remains bound, and how repeated mass-loss shapes the circumstellar environment. 

\section{Recent results}

\subsection{Classical novae}
%

The explosion of the nova V906 Carinae (V906\,Car) in 2018 provided one of the clearest demonstrations 
that even classical novae can be powerful sources of particle acceleration \citep{Aydi2020}. 
Coordinated multiwavelength observations revealed that rapid optical flares were contemporaneous 
with high-energy emission ($\gamma$-rays and hard X-rays), 
showing that a significant fraction of the optical light was produced by shock heating rather 
than only thermal radiation from the expanding shell. 
NuSTAR and XMM-Newton data confirmed the presence of hot shocked plasma, 
while follow-up radio monitoring detected clear evidence of non-thermal (synchrotron) emission 
in the immediate aftermath of the outburst and showed evolving structures consistent with colliding flows. 
Spectroscopic and infrared observations reinforced the picture that early, 
strong shocks play a central role in the radiative output and dust formation history of this nova. 
Overall, V906\,Car established a direct observational link between shocks, particle acceleration, 
and multi-band emission in classical novae \citep{Aydi2020, Sokolovsky2020}.

\begin{figure}[h]
    \centering
      \includegraphics[width=1.0\columnwidth]{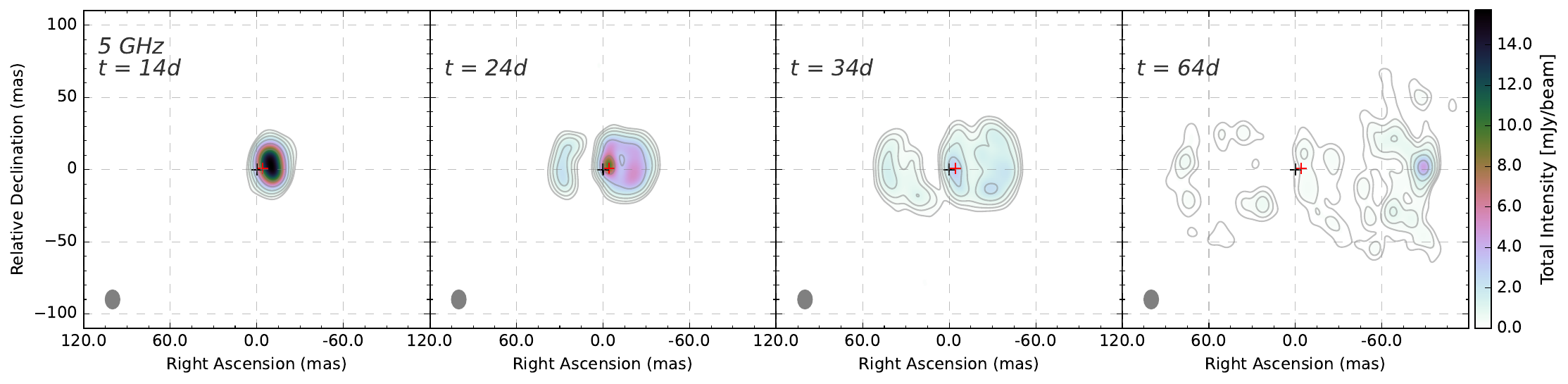}
    \caption{Natural-weighted RS\,Oph total intensity images obtained with the EVN at 5\,GHz during 2021. The observing epoch, measured in days following the outburst of 2021 August 08, appears in the top-left corner of each image. All images are convolved with a beam of 9.8 mas $\times$ 12.8 mas at 0$^{\circ}$, with the color scale and the overlaid contours representing the total intensity emission. Adapted from \citep{Lico2024}.}
    \label{fig:rsoph_lico2024}
\end{figure}

V5668 Sagittarii (V5668 Sgr), is one of the most luminous fast novae of the last decade, and was extensively observed across the electromagnetic spectrum after the 2015 eruption. Multi-band radio observations using facilities like the Atacama Large Millimeter Array (ALMA), the Jansky Very Large Array (JVLA) and the Australia Telescope Compact Array (ATCA) successfully traced the rapid evolution of its radio flux from the earliest stages, indicating that even in these fast classical novae, the expansion often deviates significantly from spherical symmetry \citep{2018MNRAS.480L..54D,Abraham2024}.



\subsection{Symbiotic novae} 
The 2021 outburst of the recurrent symbiotic nova RS Ophiuchi (RS\,Oph) provided an exceptional opportunity to study the physics of nova eruptions through high-resolution radio VLBI imaging, offering the most detailed view to date of bipolar ejecta interacting with a dense RG wind. Multi-epoch European VLBI Network (EVN) + e-MERLIN observations between 14 and 65 days after the outburst, at 1.6 and 5\,GHz (Fig.\ref{fig:rsoph_lico2024}), traced the expansion geometry of the remnant and enabled the construction of a comprehensive 3D model that connects the spatial origin of the hard and supersoft X-rays, thermal and non-thermal radio emission, and the permitted and forbidden optical/UV lines \citep{Munari2022, Lico2024}.
The radio images revealed asymmetry between the approaching and receding lobes, with the latter appearing fainter and partially absorbed in the earliest epochs. This asymmetry is naturally explained by the presence of a dense equatorial structure, identified as a Density Enhancement on the Orbital Plane (DEOP). This indicates that in addition to a massive accretion disk around the WD, the WD's gravity channels most of the RG's lost mass towards the orbital plane, creating the observed density enhancement. Consequently, the combined effect of the WD accretion disk and the DEOP confines the ejecta, forcing it to expand primarily in the bipolar structure perpendicular to the orbital plane. The density of the DEOP, as well as its degree of ionization, was found to decline from $\sim 1 \times 10^7$ cm$^{-3}$ near the binary to $\sim 9 \times 10^5$ cm$^{-3}$ at a few hundred astronomical units.
The ejecta were found to propagate at a speed of $8150$ km s$^{-1}$, with no evidence of significant deceleration, suggesting that the density gradient perpendicular to the orbital plane is very steep, and that the bulk of the ejecta's deceleration could have occurred within the initial two weeks after the eruption.
These VLBI studies provide one of the most complete, spatially resolved views yet obtained of shock propagation and bipolar outflow formation in a symbiotic nova system.


Among symbiotic novae, V407\,Cygni (V407\,Cyg) stands out as the first nova ever detected in
GeV $\gamma$-rays, following its 2010 eruption, providing clear evidence 
that shocks in nova ejecta can accelerate particles to relativistic energies \citep{Abdo2010}. 
The $\gamma$-ray emission observed by \textit{Fermi}-LAT originated as the nova ejecta impacted 
the dense wind of the Mira-type companion, generating strong shocks and non-thermal radiation while ionizing 
the surrounding circumstellar environment. 
A comparison with the recurrent symbiotic nova RS\,Oph highlights a striking difference in ejecta deceleration dynamics. 
In RS\,Oph, the bipolar lobes could have slowed significantly within the first two weeks of the eruption, whereas in V407\,Cyg the deceleration persisted for several months. This extended deceleration reflects the system's larger orbital separation, which limits the white dwarf's ability to deflect the Mira wind away from polar directions \citep{Munari2025a}.
High-resolution radio VLBI imaging also uncovered an unexpected radio-emitting feature moving radially outward along 
the same axis as the bipolar outflows, with a projected velocity of about 700 km s$^{-1}$. 
Extrapolating its trajectory places its ejection roughly 6.5 years before the 2010 eruption, 
during a phase of enhanced accretion activity, likely representing a polar blob of material expelled from the WD 
in a prior mass-ejection episode \citep{Giroletti2020}.
Such pre-eruption ejections are consistent with behavior observed in other symbiotic systems, 
most notably T Coronae Borealis (T\,CrB), which experienced a similarly super-active accretion episode prior to its 1946 outburst 
and is now approaching another anticipated eruption \citep{Munari2025b}.

\subsection{Cataclysmic variables - dwarf novae}


During the inter-eruption intervals, the classical novae build mass in their disk at a rate faster than the disk can smoothly transfer to the WD; upon reaching a threshold in mass, the disk becomes unstable and discharges most of its mass onto the WD. The potential energy released in the process powers a surge in luminosity, known as a dwarf nova outburst. Such outbursts do not involve nuclear burning and therefore are far less
energetic than those of classical novae, recur on much shorter timescales, from weeks to years, and do not eject mass into the circumbinary space \citep{2020AdSpR..66.1004H,2024A&A...689A.354J}.
In a recent radio survey using the South African MeerKAT radio telescope, \citet{Kersten2025} report the detection of 
three dwarf novae (IP Pegasi, V426 Ophiuchi, and RU Pegasi) during optical outbursts in L-band ($\sim1.28$\,GHz). 
Although several thousands of dwarf-nova systems are currently known, these are among the very few ones ever clearly detected in radio, raising the confirmed radio-emitting dwarf-nova population to about ten systems. These new detections show that dwarf novae, though intrinsically faint in the radio band, can produce detectable emission during outbursts. The emission mechanisms remain uncertain, with synchrotron emission from a jet or coronal emission from the donor star discussed as possibilities. Thanks to its superb $\mu$Jy sensitivity, the SKAO will transform this field by enabling the detection and monitoring of dozens to hundreds of dwarf novae throughout the Galaxy.

\section{The role of the SKAO in novae studies}\label{sec:role_of_SKA}

The SKAO will enable a step-change in novae science. With its unprecedented sensitivity, wide-field and VLBI imaging capabilities, it will provide the tools necessary to move beyond the study of a handful of bright, nearby novae to a comprehensive understanding of the entire Galactic nova population and beyond. 
The most significant impact for novae studies will come from SKA-Mid, that covers the 0.35 - 15.4\,GHz frequency range. The higher frequency bands of SKA-Mid, particularly band 5b at 8.3 - 15.4\,GHz where both the free-free and synchrotron emission opacity decreases, are crucial for resolving the compact structures in the early phases that follow the nova outburst. Moreover, the $\mu$Jy sensitivity of these bands will enable the characterization of the fine structure of the ejecta as well as the detection of novae from greater distances than ever before \citep{Obrien2015}.

To date, approximately 400 classical novae have been identified in the Milky Way through optical observations. In contrast, detections at radio wavelengths remain comparatively rare. Despite decades of monitoring, only a few dozen novae have been observed in the radio band. A comprehensive compilation by \citet{2021ApJS..257...49C} reports 36 classical novae with confirmed radio detections, gathered from more than fifty years of observations published in the literature. This disparity reflects both the observational challenges inherent to radio follow-ups, such as limited sensitivity and cadence, and the intrinsically diverse radio behavior of novae, which depends strongly on ejecta mass, density, and geometry. A further bias derives from the northern location of observing facilities like the JVLA, while the classical novae are primarily a Galactic Bulge population and therefore mostly located at low southern declinations.
One of the SKAO's most impactful contributions will be population-scale science. With its sensitivity, field of view and flexible cadence, the SKAO surveys can detect the radio counterparts of most Galactic novae, not only the northern and/or the brightest nearby ones, enabling a near 1:1 radio-to-optical detection rate, effectively capturing the 'faint end' of the population that is currently missed \citep{Obrien2015}. This will strongly reduce biases in ejecta mass and energetics estimates currently imposed by incomplete radio follow-up. By surveying the Galactic plane and targeting known optical transients, the SKAO will build large, homogeneous radio light-curve samples and produce the first statistically complete radio nova population. Frequent, broadband monitoring over months to years will map the variety of radio spectral evolution, characterizing both thermal and non-thermal emission, tracing shock lifecycles in a population sense, and investigating the variability, properties of mass accretion and ejection, nuclear burning, and the outflows of hundreds of novae. The multi-epoch data will provide a wealth of information on the evolution of nova light curves, enabling a more accurate determination of ejecta masses and providing a better understanding of the different physical processes at play. Beyond the outburst phase, the $\mu$Jy level sensitivity of the SKAO will allow to probe the quiescent phase of novae, characterizing the faint, steady radio emission, likely originating from the underlying binary's wind or low-level accretion processes. These observations will provide critical constraints on the mass-transfer efficiency, the nature of the accretion disk, as well as the presence, role, and intensity of the magnetic fields that shape the environment of the progenitor binary. 

Another transformative advance that the SKAO will bring to nova studies is its ability to phase the SKA-Mid dishes into a single, highly sensitive element for VLBI. When linked with existing arrays such as the EVN and global VLBI networks, the phased SKA-Mid will deliver sub-milliarcsecond resolution across a wide 1.4--15 GHz frequency range \citep{Bempong-Manful01.2026.SKA}. This configuration will significantly enhance uv-coverage, particularly in the southern hemisphere where VLBI facilities are limited, and enable ultra-sensitive, high-resolution studies of novae, including those in the Galactic Centre and Magellanic Clouds. As demonstrated by \citet{Obrien2015}, the SKAO's sensitivity in the AA* configuration will make it feasible to monitor the brightest novae in the Large Magellanic Cloud (LMC), providing a benchmark population at known distance and metallicity, both key parameters for testing how composition influences thermonuclear burning and mass ejection. Phased SKA-VLBI will deliver the sharpest radio images ever obtained of nova eruptions, and it will be crucial for investigating the ejecta geometry, shock evolution, and interactions with the circumstellar medium with unprecedented detail. A recent successful experiment between the EVN and the phased MeerKAT array in South Africa, an SKA-Mid precursor, has already validated the substantial improvements in sensitivity, uv-coverage, and resolution achieved when large southern arrays join global VLBI efforts\footnote{\tiny{\url{https://www.jive.eu/news/earth-sized-radio-observatory-just-got-better-south-africas-meerkat-telescope-joins-forces}}}. 
The future addition of African VLBI Network (AVN) stations will further enhance these capabilities, enabling the mapping of novae shock fronts on mas and sub-mas scales. Moreover, the unprecedented $\mu$as sensitivity of SKA-VLBI will allow searches for faint emission from remnants of previous outburst phases, such as that seen in V407 Cygni \citep{Giroletti2020}, potentially linked to super-active accretion phases that precede nova explosions and record the long-term history of mass ejection and wind shaping.

The full scientific potential of the SKAO will be realized through synergies with other major astronomical facilities. The SKAO's radio observations can be combined with millimeter and submillimeter wavelength data from ALMA, optical/IR photometric and spectroscopic observations by the Vera C. Rubin Observatory, X-ray observations from Swift and Athena when available, as well as high- and very-high-energy $\gamma$-ray data from space-borne and ground-based detectors, including \textit{Fermi}-LAT and the Cherenkov Telescope Array Observatory (CTAO) \citep{Castignani01.2026.SKA}. This multiwavelength approach will provide a holistic view of novae, from the initial outburst to the long-term evolution of their remnants, from the cool, dusty ejecta to the high-energy processes that accelerate particles to near the speed of light.
To ensure efficient multiwavelength coordination, the SKAO is designed with a Rapid Response Mode (RRM) that provides a level of operational flexibility beyond that of traditional arrays \citep{GemmaAnderson01.2026.SKA}. This will enable near-real-time transient alerts, autonomous triggering, and sub-arraying capabilities, allowing a portion of the array to be instantly diverted to a transient target without disrupting ongoing primary observations.

The ability to conduct quasi-simultaneous multi-band observations will also be a major asset. By observing a nova at multiple frequencies at the same time, we can perform detailed spectral studies of the ejecta. This will enable us to investigate the particle energy distribution and disentangle the different emission mechanisms at play, such as thermal bremsstrahlung and non-thermal synchrotron emission, providing a more complete picture of the physical processes within the nova shell.


\section{Concluding remarks}
The SKAO will fundamentally transform our understanding of novae, enabling a shift from the study of individual, well-known systems to a comprehensive, statistically complete census of the Galactic nova population. This will help addressing far-reaching questions like the role played by recurrent novae as progenitors of type Ia supernovae, which are key tools in the cosmic distance ladder and in probing the expansion and acceleration of the Universe.
Its unparalleled sensitivity and the possibility to use it as a phased array in global VLBI experiments will allow us to `film' the evolution of nova explosions and reveal the small-scale physics of shocks, particle acceleration and ejecta shaping with an unprecedented level of detail.
Working together with other observatories, the SKAO will reveal the complete multiwavelength picture of these thermonuclear eruptions across different environments and metallicities, answering long-standing questions while almost certainly leading to surprising new discoveries.

{\bf Acknowledgements}:
We thank the anonymous reviewer for the constructive and detailed comments, which have significantly improved the paper. BM acknowledges financial support from the State Agency for Research of the Spanish Ministry of Science and Innovation, and FEDER, UE, under grant PID2022-136828NB-C41/MICIU/AEI/10.13039/501100011033, and from the Advanced Grant from the European Research Council (ERC) under the European Union’s Horizon 2020 research and innovation programme (`EuroFlash’; Grant agreement No. 101098079). AAC acknowledges financial support from grants PID2022-136828NB-C41, and PID2022-138172NB-C43 funded by MCIN/AEI/10.13039/501100011033/FEDER, UE; the Institute of Cosmos Sciences University of Barcelona (ICCUB, Unidad de Excelencia “María de Maeztu”) through grant CEX2024-001451-M; the European Union NextGenerationEU(PRTR-C17.I1) and by Generalitat de Catalunya.

\bibliographystyle{abbrvnat-maxbibnames4}
\bibliography{chapter}

\end{document}